\documentclass[aps,prb,twocolumn,showpacs,amsmath,amssymb,superscriptaddress]{revtex4-2}
\usepackage{graphicx}
\usepackage{CJK}
\usepackage{color}
\usepackage[colorlinks,bookmarks=false,citecolor=blue,linkcolor=red,urlcolor=blue]{hyperref}
\usepackage{multirow}
\usepackage{ulem}
\usepackage{tabularx}
\usepackage{diagbox}
\usepackage[nounderscore]{syntax}
\graphicspath{{fig/}}

\newcommand{\be}{\begin{equation}}
\newcommand{\ee}{\end{equation}}

\newcommand{\trace}{{\rm Tr}}

\begin{document}
\begin{CJK*}{UTF8}{gbsn}
\title{Improved scaling of the entanglement entropy of quantum antiferromagnetic Heisenberg systems}

\author{Zehui Deng}
\affiliation{Beijing Computational Science Research Center, Beijing 100193, China}
\affiliation{Nanchang Normal University, Nanchang 330032, China}
\author{Lu Liu}
\affiliation{School of Physics, Beijing Institute of Technology, Beijing 100081, China}
\author{Wenan Guo}
\email{waguo@bnu.edu.cn}
\affiliation{Department of Physics, Beijing Normal University, Beijing 100875, China}
\affiliation{Key Laboratory of Multiscale Spin Physics, Ministry of Education, Beijing 100875, China}
\affiliation{Beijing Computational Science Research Center, Beijing 100193, China}
\author{H.Q. Lin}
\affiliation{Beijing Computational Science Research Center, Beijing 100193, China}
\affiliation{Department of Physics, Zhejiang University, Hangzhou 310027, China}
\date{\today}

\begin{abstract}
In this paper, we derive corrections to the subleading logarithmic term of the entanglement entropy in systems with spontaneous broken continuous symmetry.
Using quantum Monte Carlo simulations, we show that the improved scaling formula leads to 
much better estimations of the number of Goldstone modes in the two-dimensional
square lattice spin-1/2 Heisenberg model and bilayer spin-1/2 Heisenberg model in systems of rather small sizes, compared with previous results. 
In addition, the universal geometry-dependent finite constant in the entanglement entropy scaling is also obtained in good agreement with the theoretical value.
\end{abstract}

\maketitle

\section{Introduction}

Entanglement entropy is a valuable probe of non-local correlations for quantum systems. Thanks to conformal field theories,
the scaling of the entanglement entropy of one-dimensional critical systems is known \cite{Calabrese2004}. Our knowledge of 
entanglement entropy in two and higher dimensions is far less complete. However, one expects the leading contribution to the 
entanglement entropy scales as the area of the subsystem boundary for both critical and non-critical systems. For critical systems, a subleading 
universal but geometry-dependent constant contributes. 
In addition, there are extra subleading universal logarithmic contributions in two dimensions if the boundary has sharp corners and in three dimensions 
if the boundary is curved. 

Inspired by the quantum Monte Carlo (QMC) simulations of Kallin {\it et al.} \cite{Kallin-EE} and spin-wave calculations of Song {et al.} \cite{ee-spinwave2011},
Metlitski and Grover proposed the following scaling behavior of entanglement entropy $S$ of systems with spontaneously broken continuous 
symmetry from O($N$) to O($N-1$) \cite{Metlitski-EE} 
\begin{equation}
    S=a L^{d-1}+b \log(\frac{\rho_{s}}{c}L^{d-1})+\gamma_{\rm ord},
    \label{fssS2}
\end{equation}
where, in addition to the leading area law contribution, which scales as the area of the subsystem boundary with  a 
nonuniversal proportional constant $a$, and a universal geometry-dependent finite constant $\gamma_{\rm ord}$, 
a subleading logarithmic correction is present with $b=N_G/2$, even when the subsystem boundary contains no corners or is not curved.
Here, $N_G$ is the number of Goldstone modes, $d$ is the spatial dimension,  $c$ is the spin-wave velocity, and $\rho_{s}$ is the spin stiffness.
This formula also applies to the R\'enyi entanglement entropy $S_n$ with $a$ and $\gamma_{\rm ord}$ depending on the replica index $n$, but $b$ 
remaining unchanged. 

For the two-dimensional (2D) model with U(1) continuous symmetry, Kulchytskyy {\it et al.} \cite{Kulchytskyy}  confirmed the presence of the logarithmic term
and verified the prefactor $N_{G}=1$ as expected through QMC calculations of the second R\'enyi entropy on the 2D spin-$1/2$ XY model.
By including a $1/L$ correction to the scaling of $S_2(L)$ and assuming the theoretical value $N_G=1$, they also extracted $\gamma_{\rm ord}$,
which is found in good agreement with the analytical prediction of Metlitski and Grover \cite{Metlitski-EE} as well as the large-$S$ prediction 
given by Laflorencie {\it et al.} \cite{Laflorencie2015}. 
However, for 2D antiferromagnetic (AF) Heisenberg models with SU(2) continuous symmetry, although the logarithmic term is found by using QMC simulations,  
the estimated prefactor $N_G/2$ is in the range $0.5$ to $0.8$,
which is much smaller than the predicted value $N_G/2=1$, with the difference much larger than the statistical errors \cite{Kallin-EE, Helmes-Wessel, 
Humeniuk2012, LAFLORENCIE2016}.

The deviation of the prefactor from the expected $N_G/2=1$  was explained by the assumption that asymptotic behavior will be accessible only for 
subsystems that extend well beyond the correlation length scale \cite{Helmes-Wessel}. 
However, with the help of the recently developed algorithm based on the nonequilibrium work \cite{Alba-entropy, D'Emidio}, 
D'Emidio \cite{D'Emidio} was able to calculate the 
second R\'enyi entanglement entropy for unprecedented system sizes  up to $192\times 96$ for 
the square lattice spin-1/2 Heisenberg model. When fitting to the scaling form (\ref{fssS2}), the 
deviation to the expected value is still much larger than the error bars.  Discarding small system sizes up to 40, the deviation becomes smaller than
two error bars, with the error bar of $N_G/2$ in the order of 0.1. 
He then concluded that the convergence to the expected $N_G=2$ is very slow.
Only when magnetic order is enhanced  by a ferromagnetic next-nearest neighbor interaction, the number of Goldstone modes can be accurately extracted by fitting the scaling form (\ref{fssS2}) to entanglement entropy data with the accuracy $1 \%$\cite{D'Emidio}. 
Indeed, using a similar nonequilibrium work based algorithm, Zhao {\it et al.} \cite{zhao2022measuring} reached size up to $160\times 80$ and obtained $N_G/2=1.00(9)$.

The logarithmic term in the scaling behavior Eq. (\ref{fssS2}) originates from the interplay of Goldstone modes and restoration of symmetry in a 
finite volume \cite{Metlitski-EE}.  The simple model reproducing the logarithmic term contains two coupled quantum rotors,
where the excited energies of the rotors are assumed to be determined by the transverse susceptibility in the thermodynamic limit, and the coupling is 
taken as the spin stiffness in the thermodynamic limit multiplied by $L^{d-2}$ with $d$ the dimensionality.
In this paper, we will show that, to better describe the scaling behavior of the entanglement entropy, it is crucial to use the finite-size spin stiffness instead of the spin stiffness in the thermodynamic limit
to describe the coupling of the two rotors and go beyond the leading order excitation energies of the rotors, which is 
used by Metlitski and Grover \cite{Metlitski-EE}.  By adopting the chiral perturbation theory \cite{chiral-perturbation} of SU(2) antiferromagnets, we
include a correction term to the excitation energies of the models.
This way, we will derive a modified scaling formula of the entanglement entropy compared to the scaling form (\ref{fssS2}).
We then 
revisit the square lattice AF Heisenberg model and the bilayer AF Heisenberg model at several ratios of the exchange couplings. With the help of 
our improved finite-size scaling formula of $S_n$, we show that the coefficient of 
the subleading logarithmic term converges to the expected $N_G/2$ within the accuracy 1 $\%$ for the square lattice Heisenberg model using 
modest system sizes up to $40 \times 40$, and within the accuracy of $5 \%$ for several values of coupling ratios of the bilayer Heisenberg model 
using system sizes up to $24 \times 24$.
Our modified scaling formula also explains why the quantum XY model follows the original formula Eq. (\ref{fssS2}) confirmed by Kulchytskyy {\it et al.} 
\cite{Kulchytskyy}.

The paper is organized as follows: Sec. \ref{theory} we derive the modified scaling formula of the entanglement entropy with corrections due to 
finite-size effects of spin stiffness and excitation energies beyond leading order. 
Sec. \ref{method} describes the numerical method and physical quantities used in this work.
In Sec. \ref{2DHeisenberg} and \ref{blH}, we present numerical results of the 2D AF Heisenberg model and the bilayer AF Heisenberg model.  
We analyze the finite-size results with our improved scaling formula and extract $N_G$ with unprecedented accuracy using only modest system sizes. We 
conclude in Sec. \ref{sec_conclusion}.

\section{Modification of the logarithmic term}
\label{theory}
The 2D Heisenberg antiferromagnet is described by the nonlinear sigma model, with its low-energy properties in the thermodynamic limit determined by
the spin stiffness $\rho_s$, the spin-wave velocity $c$, and the staggered magnetization $m_s$. This description also applies to the spin-1/2 AF 
Heisenberg model on the square-lattice bilayer in its N\'eel phase. The nonlinear sigma model can be generalized to $N$-component vectors with 
O($N$) symmetry and  the quantum AF Heisenberg model corresponds to the $N=3$ case.

The ground state of the quantum O($N$) model on $d\ge 2$ dimensions is known to spontaneously break the O($N$) symmetry to an O($N-1$) symmetry,
which is infinitely degenerate, labeled by the N\'eel order parameter, in the thermodynamic limit. 
However, for a system with a finite size, the system has a unique ground state and a tower of excited states described by quantum rotors, with 
excitation energies 
\begin{equation}
    E_{S}-E_{0}=\frac{S(S+N-2)}{2\chi_\perp L^d},
    \label{energy1}
\end{equation}
where $S$ is the angular momentum number of the total spin ${\bf S}$ of the system, $\chi_\perp=\rho_s/c^2$ is the transverse susceptibility in the 
thermodynamic limit, which 
is related to the uniform susceptibility $\chi_u$ in the thermodynamic limit at zero temperature through $\chi_\perp=\frac{3}{2}\chi_u$ for the $N=3$ case.
Here, $\chi_\perp L^d$ can be considered as the effective moment of inertia of a rotor. We can define $I = \chi_\perp$ as the effective inertia moment density.


The chiral perturbation theory \cite{chiral-perturbation}
predicts scaling forms of size dependence of various quantities beyond leading order. According to the chiral perturbation theory, 
the excitation energies Eq. (\ref{energy1}) has a finite-size correction in $d=2$,  
\begin{equation}
    E_{S}-E_{0}=\frac{S(S+N-2)}{2\chi_\perp L^2}[1-\frac{(N-2)}{c \chi_\perp L}\frac{3.900265}{4\pi}+O(\frac{1}{L^2})]
    \label{energy2}
\end{equation}
with $N=3$ for Heisenberg antiferromagnets. 
Essentially, this is to replace the inertial moment density $I=\chi_\perp$ to a finite-size dependent inertia moment density
\begin{equation}
	I(L)=\chi_\perp[1+\frac{(N-2)}{c\chi_\perp L}\frac{3.900265}{4\pi}+O(\frac{1}{L^2})].
    \label{inertia}
\end{equation}

Eq. (\ref{inertia}) is the special case for $L_{x}=L_{y}=L$, with $L_{x}$ and $L_{y}$ the linear size in $x$ and $y$ direction, respectively. 
For more general cases, the inertial moment density is written as
\begin{equation}
	I(L_x, L_y)=\chi_\perp[1+\frac{(N-2)}{c\chi_\perp \sqrt{V}}\frac{3.900265}{4\pi}+O(\frac{1}{L^2})]
    \label{inertia2}
\end{equation}
with the volume of the system $V=L_{x}L_{y}$.

To capture the subleading logarithmic term in the scaling form Eq. (\ref{fssS2}), following Metlitski and Grover \cite{Metlitski-EE}, we consider the simple quantum mechanical model of two coupled quantum O($N$)
rotors ${\bf n}_A$ and ${\bf n}_B$, representing the average order parameter in subsystem $A$ and its complement $B$. 
The Hamiltonian of the model reads 
\begin{equation}
    H=\frac{{\bf S}_A^2}{2I(L) V_{A}} +\frac{{\bf S}_B^2}{2I(L)V_{B}} -J(L) {\bf n}_A \cdot {\bf n}_B,
\end{equation}
with ${\bf S}_{A,B}$ the angular momentum of each rotor. 
$V_{A}$ and $V_{B}$ denote the volumes of each subsystem. The total volume of the system is $V=L^d=V_A+V_B$, with $d$ the dimensionality 
of the system.   
We set $J(L) =\rho_s(L)L^{d-2}$ to reflect the order-parameter stiffness of the system with $\rho_s(L)$ the spin stiffness in a finite system.   
Comparing with the model of Metlitski and Grover \cite{Metlitski-EE},  we have used finite-size dependent inertia moment density $I(L)$ replacing $\chi_\perp$ 
and finite-size dependent $J(L)$ replacing $J=\rho_s L^{d-2}$.  Note $\rho_s$ is the spin stiffness in the thermodynamic limit, but $\rho_s(L)$ 
is the finite-size value of the spin stiffness. 

The model can be solved in the same way presented in \cite{Metlitski-EE}: By introducing the average and relative coordinates
${\bf n}$ and $\delta {\bf n}$, the system is decoupled into a quantum rotor with total angular momentum ${\bf S}={\bf S_A}+{\bf S_B}$ 
and total moment of inertial $I(L) V$ and an $N-1$ dimensional harmonic oscillator with the frequency 
\begin{equation}
    \omega=\sqrt{\frac{\rho_s(L) L^{d-2}}{I(L)V_r}}\sim \sqrt{\frac{\rho_s(L)}{I(L)}}\frac{1}{L}
    \label{dispersion}
\end{equation}
where $V_r=V_{A}V_{B}/(V_{A}+V_{B})$ is the reduced volume. Then, the logarithmic diverging R\'enyi entanglement entropy is obtained straightforwardly. 
As a result of including finite-size dependent parameters $I(L)$ and $\rho_s(L)$, Eq. (\ref{fssS2}) becomes 
\begin{equation}
	S_{n}(L) =a L^{d-1} + \frac{N_G}{2} \log(I(L)^{1/2}\rho_s(L)^{1/2} L^{d-1}) +\gamma_{\rm ord},
	\label{newS2}
\end{equation}
with a modified subleading logarithmic term,
which reduces to the logarithmic term in Eq. (\ref{fssS2}) when replacing $\rho_s(L)$ to $\rho_s$ and $I(L)$ to $\rho_s/c^2$.

For the 2D quantum XY model, since $N=2$, $I(L)=\chi_\perp$ up to the leading order $1/L$ according to Eq. (\ref{inertia}). Expanding $\rho_s(L)$ to 
$\rho_s + b/L$, we can write Eq. (\ref{newS2}) as
\begin{equation}
	S_{n}(L) =a L^{d-1} + \frac{N_G}{2} \log(\frac{\rho_s}{c} L^{d-1}) + \frac{N_{G}}{4}\frac{b}{\rho_s L} + \gamma_{\rm ord}.
	\label{newS2XY}
\end{equation}
This explains why Kulchytskyy {\it et al.} confirmed the presence of the logarithmic term with the prefator $N_G=1$ and extracted $\gamma_{\rm ord}$
in good agreement with analytical prediction \cite{Metlitski-EE} by including a $1/L$ correction to the scaling of $S_2(L)$ Eq. (\ref{fssS2}).

The bilayer AF Heisenberg model can also be described by the nonlinear sigma model if 
the spin stiffness and transverse susceptibility are defined in a unit of the unit cell of the model. Then, the finite-size dependent effective inertia moment density 
Eq.(\ref{inertia}) also applies to the bilayer Heisenberg model. Therefore, the modified scaling formula of $S_2(L)$, Eq. (\ref{newS2}), should hold.

\section{Numerical results}
\subsection{Qunantum Monte Carlo Methods}
\label{method}

The R\'enyi entanglement entropy is defined as
\begin{equation}
    S_{n}(A)=\frac{1}{1-n}\ln{\trace [\rho_{A}^{n}]},
    \label{Sn_def}
\end{equation}
where $n$ is the R\'enyi index( $n=2$ in our work) and $\rho_{A}=\trace_{\bar{A}}{\rho}$ is the reduced density matrix of a subsystem $A$ 
with $\bar{A}$ its complement. $\rho=e^{-\beta H}/Z$ is the density operator with $Z=\trace e^{-\beta H}$ the partition function. $\beta \to \infty$ is 
the inverse temperature to probe only the properties of the ground states. 

With the help of the replica trick \cite{Calabrese2004}, the R\'enyi entanglement entropy 
$S_n(A)$ can be expressed as the ratio of free energies, which can be calculated much more efficiently by  
using the algorithm developed recently \cite{Alba-entropy,D'Emidio} with the help of the nonequilibrium work relations \cite{Jarzynski,Crooks,bulgarelli2023entanglement}.
In this work, we make use of the version for the projector quantum Monte Carlo method (PQMC) \cite{sandvik-pqmc, pqmc-loopupdate} to extract 
the R\'enyi entanglement entropy $S_2$ \cite{D'Emidio}.  

We set the projection power $m=20 N_s$,  which is large enough to probe for the ground state properties 
as shown in the supplemental material of Ref. \cite{D'Emidio} for the Heisenberg model and appendix \ref{S2mconverge} of the current paper for the bilayer 
Heisenberg model at various $g$.  
Here, $N_s$ denotes the total number of spins of the quantum systems. In particular, for the square lattice Heisenberg model, $N_s=L_x \times L_y$, while 
in the case of the bilayer Heisenberg model, $N_s=2 L_x \times L_y$, with $L_x=L_y=L$.
In our simulations, we consider bipartite  the toroidal lattice into two equally sized cylindrical strips of 
size $N_{A}=L/2\times L$ and study the R\'enyi entanglement entropy of one subregion. 
In the simulations of the square lattice Heisenberg model, we compute 1000 nonequilibrium work realizations for system sizes ranging from $L=8$ to 36
and 2000 nonequilibrium work realizations for $L=40$.
Each work realization consists of $N_{A}\times 10,000$ nonequilibrium time steps. 
For the bilayer Heisenberg model, since $N_{A}$ includes twice as spins as that of in the single layer model, we 
choose each nonequilibrium work realization consisting of $N_{A}\times 20,000$ nonequilibrium time steps.

To calculate the spin stiffness and the susceptibilities, we apply the stochastic series expansion (SSE) QMC method with the loop update algorithm \cite{Sandvik1997, Sandvik-review}.

The spin stiffness is defined as the free energy increasing {\it per unit cell} due to the presence of a twist field,
\begin{equation}
\rho_{s}=\frac{3}{2 N}\frac{\partial^{2} F(\phi)}{\partial\phi^{2}},
\end{equation}
where $F(\phi)$ is the free energy in the presence of a twist field $\Phi$. 
Here $N=L_x\times L_y$ is the number of unit cells. 
In Monte Carlo simulations, the spin stiffness $\rho_s(L)$ is calculated through the fluctuations of the winding number of spin transporting
\begin{equation}
    \rho_{s}=\frac{3}{4\beta N}\langle{ L_x^2 W^{2}_{x}+ L_y^2 W^{2}_{y}}\rangle,
    \label{stiffness}
\end{equation}
where the winding numbers are defined as
\begin{equation}
    W_{\alpha}=(N^{+}_{\alpha}-N^{-}_{\alpha})/L_\alpha.
\end{equation}
Here, $N^{+}_{\alpha}(N^{-}_{\alpha})$ is the total number of operators transporting spin in the positive (negative) $\alpha=x, y$ direction. Note that $N=N_s/2$ for the bilayer Heisenberg model. 

In order to calculate the uniform susceptibility $\chi_u$, we consider the wave-vector ${\bf q}$-dependent 
susceptibility $\chi({\bf q})$ \cite{Wang-dblH}, which is the Fourier transform of the static spin-spin susceptibility in real space $\chi(k_\sigma,l_{\sigma^\prime})$
\begin{equation}
    \chi({\bf q})=\frac{1}{N}\sum_{k,l}\sum_{\sigma,\sigma^{\prime}} e^{i {\bf q}\cdot ({\bf r}_k-{\bf r}_l}) \chi(k_{\sigma},l_{\sigma^{\prime}})
    \label{chiq}
\end{equation}
with 
    \begin{equation} \chi(k_{\sigma},l_{\sigma^{\prime}})=\int_{0}^{\beta}d\tau\langle{S_{k_{\sigma}}^{z}(\tau)S_{l_{\sigma^{\prime}}}^{z}(0)}\rangle,
\end{equation}
which is obtained using standard SSE simulations \cite{Sandvik1997}. 
Here, $k_{\sigma} (l_{\sigma^\prime})$ denotes the spin in $\sigma (\sigma^\prime)$-th layer in the  unit 
cell $k$ ($l$). 
For the single-layer Heisenberg model, $\sigma, \sigma^{\prime}=1$, while for the bilayer Heisenberg 
model, $\sigma,\sigma^{\prime}=1,2$. $N$ is the total number of unit cells in the system. 

The value of $\chi({\bf q})$ at the longest wavelength, ${\bf q}=(2\pi/L,0)$, is taken as the definition of the 
finite-size uniform susceptibility $\chi_u(L)=\chi(2\pi/L,0)$, which converges to $\chi_u$ when $L \to \infty$ \cite{Wang-dblH}.
Therefore, to obtain the transverse susceptibility $\chi_\perp$ of a system with broken symmetry in the thermodynamic limit, we define the 
finite-size transverse susceptibility as
\begin{equation}
    \chi_{\perp}(L)=\frac{3}{2}\chi(2\pi/L,0),
\end{equation}
which converges to $\chi_\perp$ at the limit $L\to \infty$.

Simulations of $L\times L$ systems for the Heisenberg model and $2\times L \times L$ systems for the bilayer Heisenberg model were carried out at inverse 
temperature $\beta=4L$ and $\beta=6L$, respectively. The $\beta$s we have chosen here ensure the convergence of the spin stiffness and uniform susceptibility 
to their ground state values within 
statistical errors; see appendix \ref{beta_gs} for details.

\subsection{The square lattice spin-1/2 AF Heisenberg model}
\label{2DHeisenberg}

In this section, we consider the spin-1/2 AF Heisenberg model on the square lattice with Hamiltonian 
\begin{equation}
    H=J\sum_{\langle{i,j}\rangle}{\bf S}_{i}\cdot {\bf S}_{j},
\end{equation}
where $\langle{i,j}\rangle$ are nearest neighbors on a periodic square lattice with $L^{2}$ sites and $J>0$ is the exchange interaction.

\begin{figure}[h]
     \centering
     \includegraphics[width=0.5\textwidth]{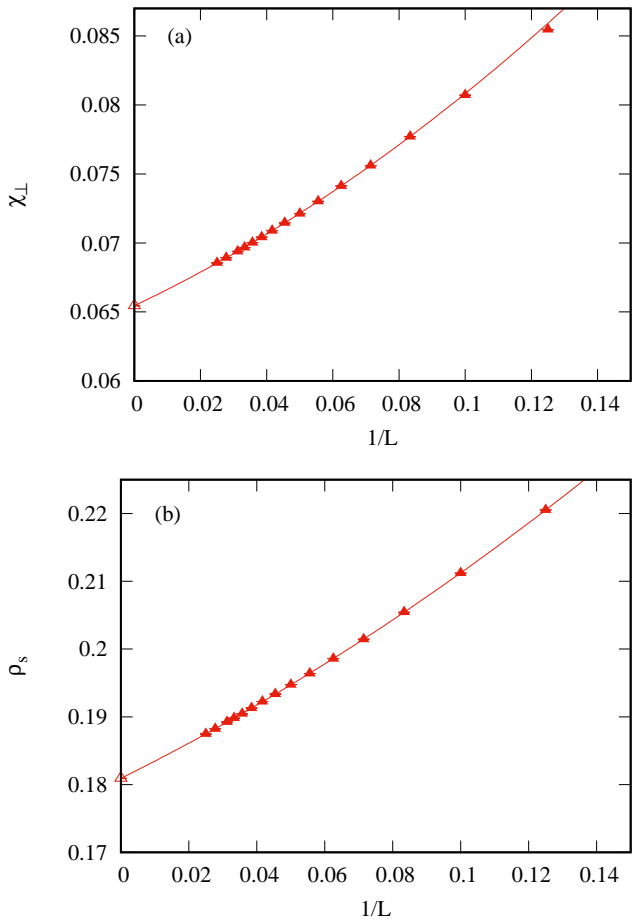}
     \caption{ $\chi_\perp(L)$ and $\rho_s(L)$ of the AF Heisenberg model on square lattice. (a) $\chi_\perp(L)$ versus $1/L$. The solid line is the fit  according to Eq. (\ref{chi_fss}) 
     up to the second order of $1/L$. (b) $\rho_s(L)$ versus $1/L$. The solid line is the fit according to Eq. (\ref{rhos_fss}) up to the second order of $1/L$. Error bars are much smaller than the symbols. The estimated $\chi_\perp$ and $\rho_s$ are also shown.}
     \label{chirhos_H}
 \end{figure}

Figure \ref{chirhos_H} shows QMC results of the transverse susceptibility $\chi_\perp(L)$ and spin stiffness $\rho_s(L)$ versus system size $L$.

We analyze $\chi_{\perp}(L)$ to obtain the thermodynamic limit value $\chi_\perp$, using the following expansion \cite{Sandvik1997} 
\begin{equation}
    \chi_{\perp}(L)=\chi_{\perp}(1+\frac{b_{1}}{L}+\frac{b_{2}}{L^2}+\cdots),
    \label{chi_fss}
\end{equation}
where $b_i$ are constants, and $\chi_\perp$ is the susceptibility at the thermodynamic limit $L \to \infty$. 

In this work, data analysis is based on the nonlinear least-square fitting with the Levenberg-Marquardt method \cite{PYoung}. The error bars on the fit 
parameters are obtained, as well as the value of the fit parameters. $\chi_r^2\equiv \chi^2/N_{\rm DOF}$ with $N_{\rm DOF}$ the number of degrees of freedom 
shows the goodness of the fit.  For $N_{\rm DOF} \gg 1$, $\chi_r^2 \approx 1$ is expected for a statistically sound fit.  
The P-value of the fit,  which describes the distribution of $\chi^2$,  is useful when $N_{\rm DOF}$ is not large: it should take a value in the range of $0.05$ to $0.95$ for a statistically sound fit.

Fitting our QMC data $\chi_\perp(L)$ according to Eq. (\ref{chi_fss}) up to the second order of $1/L$, we 
obtain $\chi_\perp=0.06545(3)$ with $b_{1}=1.74(2), b_{2}=6.1(1)$ in a statistically sound fit with $\chi_r^2=1.38$ and P-value $0.18$.

Now we try to find out the finite-size behavior of $\rho_s(L)$, which can be written as follows \cite{Sandvik1997,Sandvik-review}
\begin{equation}
  \rho_{s}(L)=\rho_{s}(1+\frac{a_{1}}{L}+\frac{a_{2}}{L^2}+\cdots),
    \label{rhos_fss}
\end{equation}
where $\rho_s$ is the spin stiffness at the thermodynamic limit $L\to \infty$. 
Fitting Eq. (\ref{rhos_fss}) up to the second order of $1/L$ to our QMC data $\rho_s(L)$, we find $\rho_s=0.18092(3)$ in good agreement with or close to results 
in the literature \cite{Sandvik1997, FJJiang, Jiang-Wiese} and the values of $a_1=1.375(7)$ and $a_2=3.01(4)$ with $\chi_r^2=0.90$ and P-value $0.54$.

  \begin{figure}[h]
     \centering
     \includegraphics[width=0.5\textwidth]{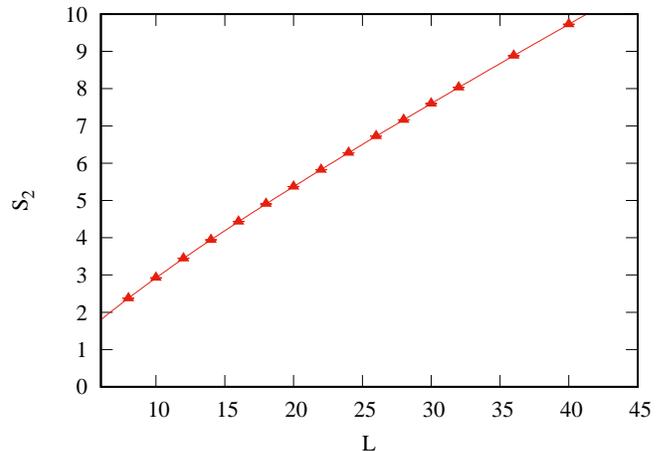}
     \caption{The second R\'enyi entropy $S_{2}(L)$ versus $L$ for the square lattice Heisenberg model with error bars much smaller than the size of the symbol. The solid line is a fit according to Eq. (\ref{newS2_1}).}
     \label{S2_H}
 \end{figure}

 With estimated $\chi_\perp$ and $\rho_s$, we obtain $c=\sqrt{\rho_s/\chi_\perp}=1.66260(7)$ which is in good agreement with reults in the 
 literature \cite{FJJiang,Jiang-Wiese,Sandvik1997}. The finite-size dependent effective inertia moment density $I(L)$ is then obtained 
 to the order O($1/L$) using Eq. (\ref{inertia}). 

 Now, we are in the position to test the modified scaling Eq. (\ref{newS2}). 
 
 Figure \ref{S2_H} shows $S_2(L)$ versus $L$, obtained using the PQMC version of the nonequilibrium work algorithm \cite{D'Emidio}. Substituting the 
fitted function Eq.~(\ref{rhos_fss}) and  $I(L)$ with fitted $\chi_\perp$ and $c$ 
 into Eq. (\ref{newS2}), we obtain the finite-size behavior of $S_2(L)$ for the 2D AF Heisenberg model: 
\begin{equation}
	\begin{split}
		&S_{n}(L)=a L^{d-1} + \frac{N_G}{2} \\
		&\times \log((\chi_\perp[1+\frac{3.900265}{4\pi c\chi_\perp L }])^{1/2}(\rho_s(1+\frac{a_1}{L}+\frac{a_2}{L^2}))^{1/2} L^{d-1})\\
		& +\gamma_{\rm ord},
	\end{split}
	\label{newS2_1}
\end{equation}
with $\chi_\perp, c, \rho_s, a_1, a_2$ found from simulation results of $\chi_\perp(L)$ and $\rho_s(L)$.

 Fitting our QMC results of $S_2(L)$ according to this formula with $a, N_G, \gamma_{\rm ord}$ unknown, we find a statistically sound fit for all $S_2(L)$. 
 However, as listed in Tab. \ref{table1}, the difference between the obtained $N_G/2$ and the theoretical value is about two statistical errors, even though the fit is statistically sound with $\chi_r^2=1.17$
 and P-value 0.30. We can then conclude that the fit has systematical errors due to ignoring higher-order corrections in Eq. (\ref{newS2}).
 We then increase the smallest size $L_{\rm min}$ to 12 in the analysis and 
 again obtain a statistically sound fit, with $N_G/2$ differing from 1 within one statistical error, as listed in Tab. \ref{table1}. Apparently, the 
 systematical error has been removed by excluding data of the smallest size $L=8$. Upon excluding even more points, the error bars on the fit parameters 
 increase rapidly; still, the fit remains statistically sound, and the extracted $N_G$ is in good agreement with the expected values statistically.
 We take $N_G/2=0.99(1)$ and $\gamma_{\rm ord}=0.78(3)$ as our final estimates. 
 The estimated $\gamma_{\rm ord}$ is in good agreement with the theoretical value \cite{Metlitski-EE}.

 \begin{ruledtabular}
 \begin{table}[thb]
  \caption{Coefficients of the area law, the logarithmic term, and the constant. }
    \begin{tabular}{c|c|c|c|c}
         $L_{\rm min}$ &  a           & $N_G/2$    &$\gamma_{\rm ord}$ &$\chi_r^2$/P-value \\   
    \hline
                8  &  0.1860(2)   & 0.991(6)   &   0.78(3)                & 1.17/0.30   \\
    \hline
                12 &  0.1861(3)   & 0.99(1)    & 0.78(3)                  &1.13/0.33 \\
    \hline
                16 &  0.1864(6)   & 0.98(2)    &  0.79(5)                 & 1.07/0.38 \\
    \hline
                20 &  0.1855(9)   & 1.01(3)    &   0.77(8)                & 1.10/0.36\\     
     \end{tabular}
     \label{table1}
 \end{table}
 \end{ruledtabular}

 Alternatively, we can fit $S_2(L)$ according to the following equation: 
\begin{equation}
	\begin{split}
		&S_{n}(L)=a L^{d-1} + \\
		&\frac{N_G}{2} \log((\chi_\perp[1+\frac{3.900265}{4\pi c\chi_\perp L }])^{1/2}(\rho_s(L))^{1/2} L^{d-1})\\
		& +\gamma_{\rm ord},
	\end{split}
	\label{newS2_2}
\end{equation}
with $\chi_\perp, c$ known and $\rho_s(L)$ reading from numerical data. The parameters to be fitted are $a, N_G$, and $\gamma_{\rm ord}$.
This leads to the same result (within one error bar) obtained above.

Conversely, if we use the scaling form Eq. (\ref{fssS2}), in which the finite-size effect of spin stiffness and inertia moment density are ignored, to fit 
the data, we would obtain $N_G/2=0.69(2)$, which is a coincidence with the value obtained in lieterature \cite{Humeniuk2012, LAFLORENCIE2016, Helmes-Wessel, 
D'Emidio}.

\subsection{Bilayer Heisenberg model}
\label{blH}

In this section, we consider the spin-$1/2$ AF Heisenberg model on the bilayer square-lattice with Hamiltonian described by the following equation
\begin{equation}
    H=J\sum_{\langle{i,j}\rangle, \sigma}{\bf S}_{i_\sigma}\cdot {\bf S}_{j_\sigma}
	+J_{\perp}\sum_{i}{\bf S}_{i_1}\cdot{\bf S}_{i_2},
    \label{dblHeisenberg}
\end{equation}
where $i$ denotes the $i$-th unit cell containing two spin-1/2 degrees of freedom and $\sigma=1, 2$ represents layers. $\langle{i,j}\rangle$ are nearest neighboring unit cells. $J$ and $J_{\perp}$ represent the intralayer and the interlayer exchange interactions, respectively. We denote the ratio of the exchange interactions as $g=J_{\perp}/J$. 

This is a basic quantum spin model that exhibits a well-characterized quantum phase transition in the ($2+1$)-D O(3) universality class at 
the critical value $g_{c}=2.5220(1)$, which separates the antiferromagnetic ordered phase from the magnetically disordered dimer spin singlet phase \cite{Wang-dblH, Liulu}.

Helmes and Wessel \cite{Helmes-Wessel} studied the scaling of the R\'enyi entanglement entropy $S_2(L)$ of this model.
They analyzed the subleading logarithmic contribution to the R\'enyi entanglement entropy scaling upon varying the interaction ratio $g$
and obtained values of $N_G/2$ between 0.7 and 0.8 for $0<g<g_c$, while for $g=0$, they obtained $N_G/2=1.35(2)$.
They attributed the deviation to the expected behavior Eq. (\ref{fssS2}) to the size of the subsystems: the asymptotic behavior is accessible only for
subsystems that extend well beyond the correlation length scale. They also tried including further subleading finite-size correction, which
scales with $1/L$, but found that such  
fitting ansatz results in significant uncertainties on the fit parameters.

In this section, we will present our results of $S_2(L)$ 
and analyse the scaling behaviors of $S_2(L)$ using our improved finite-size scaling formula Eq. (\ref{newS2}) in the 
antiferromagnetic ordered phase.

\begin{figure}[thb]
     \centering
     \includegraphics[width=0.5\textwidth]{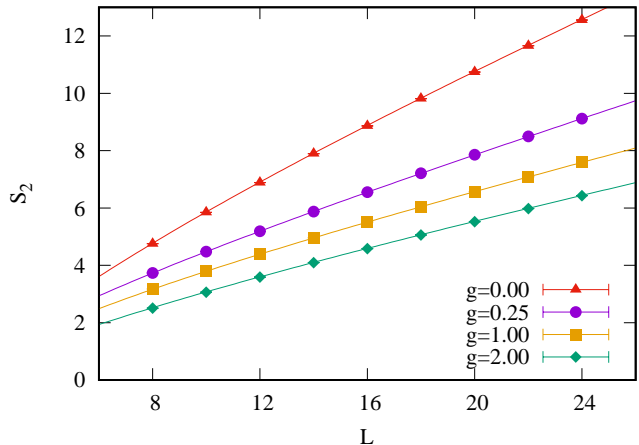}
     \caption{The second R\'enyi entropy $S_{2}(L)$ of the bilayer Heisenberg model versus $L$ at $g=0, 0.25, 1, 2$. 
     Error bars are much smaller than the sizes of symbols. The solid lines are fitting curves discussed in the text.
     }
     \label{S2_dblH}
 \end{figure}

We perform QMC simulations at several values of $g$ and calculate $S_2(L)$ up to $L=24$ 
using the PQMC version of the nonequilibrium work algorithm \cite{D'Emidio}.
Figure \ref{S2_dblH} shows $S_2(L)$ versus $L$ at $g=0$, 0.25, 1, and 2.

At $g=0$, where the symmetry is trivially enhanced to a $SU(2)\times SU(2)$ due to the decoupling of the layers,  there are four Goldstone modes existing 
in the system, i.e., $N_{G}=4$. 
 
Since the coupling between two layers is absent, $\chi_\perp$, $\rho_s$, and $c$ are the same as those of the Heisenberg model on the square lattice. 
The functions $I(L)$ and $\rho_s(L)$ are also the same as those of the Heisenberg model on the square lattice. 
Substituting these functions into Eq. (\ref{newS2}),  we fit $S_2(L)$ according to Eq. (\ref{newS2_1}).  
Using $L \le 12$ points,
we obtain a statistically sound fit for all $S_2(L)$ data with  $N_G/2=2.02(2)$, $\gamma_{\rm ord}=1.58(4)$, and $a=0.3700(9)$.  
Further excluding small-$L$ points does not dramatically change the fit parameters, though of course the error bar grows.
These results are listed in Tab. \ref{tab:blHS2}.
The estimated $N_G$ agrees with the expected value within the statistical error.
The value of $\gamma_{\rm ord}$ also coincides with the expected value $2\times 0.77$ \cite{Metlitski-EE}.
Compared with the results obtained in \cite{Helmes-Wessel}, where $N_G/2$ was found to be $1.35(4)$, our result of $N_G/2$ is much better.   

\begin{ruledtabular}
 \begin{table}[thb]
  \caption{Coefficients of the area law, the logarithmic term, and the constant at several $g$ of the bilayer Heisenberg model. }
     \begin{tabular}{c|c|c|c|c}
         g  &   $a$         & $N_G/2$     &$\gamma_{\rm ord}$ & $\chi_r^2$/P-value\\
    \hline
          0 &  0.3700(9)  & 2.02(2)     & 1.58(5)             &   0.89/0.50        \\
    \hline
       0.25 &  0.273(1)   & 1.00(3)     & 0.74(5)             &  1.60/0.16          \\
    \hline
          1 &  0.210(1)   & 1.05(3)     & 0.71(5)             &  0.69/0.63           \\
    \hline
          2 &  0.184(1)   & 1.02(3)     & 0.76(5)             &  0.45/0.81            \\
     \end{tabular}
     \label{tab:blHS2}
 \end{table}
\end{ruledtabular}

\begin{ruledtabular}
 \begin{table}[thb]
  \caption{$\chi_\perp$, $\rho_s$, and $c$ at several $g$ of the bilayer Heisenberg model. }
     \begin{tabular}{c|c|c|c}
         g &  $\chi_\perp$& $\rho_s$  & $c$      \\
    \hline
       0.25&  0.14664(6)  & 0.4223(3) & 1.69706(7)  \\
    \hline
          1&  0.12978(6)  & 0.4116(1) & 1.78088(2)    \\
    \hline
          2&  0.07144(4)  & 0.2490(2) & 1.86693(3)    \\
     \end{tabular}
     \label{tab:blH}
 \end{table}
\end{ruledtabular}

 \begin{figure}[tbh]
   \centering
    \includegraphics[width=0.5\textwidth]{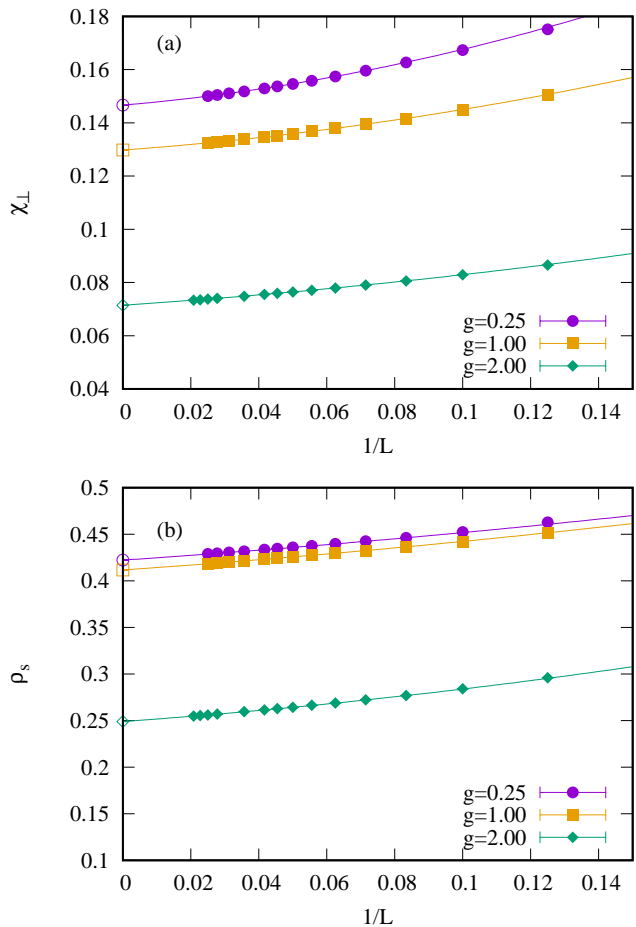}
     \caption{(a) $\chi_\perp(L)$ and (b) $\rho_s(L)$ versus $1/L$ for the bilayer Heisenberg model at $g=0.25, 1, 2$.  Error bars are much smaller than symbol 
     sizes. The solid lines are fitting curves. $\chi_\perp$ and $\rho_s$ for each $g$ are also shown.} 
      \label{chirhos_dblH}
 \end{figure}
 
 There are couplings between two layers for $g=0.25, 1,$ and $2$. To describe the bilayer AF Heisenberg model with the nonlinear sigma model, it 
 is necessary to define the spin stiffness and susceptibility in a unit of the unit cell consisting of two spins in different layers, as we do in 
 Eq. (\ref{stiffness}) and Eq. (\ref{chiq}).
$\chi_\perp(L)$ and $\rho_s(L)$ at $g=0.25, 1, 2$ are calculated up to $L=40$ using the standard SSE method. The 
results as functions of $L$ are illustrated in Fig.~\ref{chirhos_dblH}(a) and (b), respectively. We then find $\chi_\perp$ and $\rho_s$ by fitting 
finite-size data of $\chi_\perp(L)$ and $\rho_s(L)$ according to Eq. (\ref{chi_fss}) and 
 (\ref{rhos_fss}), respectively, and then calculate $c$.  The results are obtained and listed in Tab. \ref{tab:blH}. 
 The fitted parameters $b_1, b_2, a_1, a_2$ at various $g$ are also provided in Tab. \ref{tab:ab} in appendix \ref{app_ab}.

 As a result of defining the spin stiffness and susceptibility in a unit of the unit cell, 
 the finite-size dependent effective inertia moment density Eq. (\ref{inertia})
 of the nonlinear sigma model also applies to the bilayer AF Heisenberg model. Therefore, the modified scaling formula of $S_2(L)$, Eq. (\ref{newS2_1}), 
 making use of the function $I(L)$, calculated using fitted $\chi_\perp$ and $c$, and fitted function $\rho_s(L)$ for the bilayer model in Eq. (\ref{newS2}), 
 is expected to hold here. 

Fitting Eq. (\ref{newS2_1}) to $S_2(L)$, we obtain statistically sound fits of $a$, $N_G/2$, and $\gamma_{\rm ord}$ with all system sizes included for 
each $g$, as illustrated in Tab. \ref{tab:g0.25}, \ref{tab:g1}, and \ref{tab:g2} in Appendix \ref{lmin}. 
Systematical errors are present for the case $g=0.25$ according to the fit result of $N_G$. For $g=2$, the difference between fitted $N_G/2$ and the theoretical 
value is two error bars. These results are 
due to ignoring higher-order corrections in Eq. (\ref{newS2}) for small system sizes. By excluding $L=8$ point, these systematical errors are removed. 
Further excluding small-$L$ points in the fits for different $g$ does not dramatically change the fit parameters, though the error bar grows.
The best estimates are obtained with $L_{\rm min} = 10$ at each $g$. Our final estimates for $a, N_{G}/2$, and $\gamma_{\rm ord}$ are listed in 
Tab. {\ref{tab:blHS2}}.

 \section{conclusion and discussion}
 \label{sec_conclusion}
 In this paper, we have derived an improved scaling of entanglement entropy in systems with the spontaneous broken of the continuous O($N$) symmetry
 using finite-size spin stiffness and going beyond the leading order excitation energies of the rotors in the model introduced by Metlitski and Grover.
 Using QMC simulations, we have shown that our scaling formula is correct and valuable by extracting $N_G$ and $\gamma_{\rm ord}$ 
 with unprecedented accuracy in systems of rather small sizes for the 2D square 
 lattice Heisenberg model and the double layer Heisenberg model at various $g$. 
 
 In addition, from Eq. (\ref{inertia}), we know the correction to the inertial moment density $I(L)$ due to the finite-size effect can not be ignored 
 except for $O(N=2)$. Thus, for systems with spontaneously broken $O(N>2)$ symmetry, our scaling formula Eq. (\ref{newS2}) should be used to extract 
 the properties of the entanglement entropy faithfully. 

\begin{acknowledgments}
This work was supported by the National Natural Science Foundation of China under Grant No.~12175015 and No.~11734002 and the Science and Technology Foundation of Jiangxi Provincial Department of Education under Grant No.~GJJ181094 and Beijing Institute of Technology Research Fund Program for Young Scholars.
The authors acknowledge the support of the Super Computing Center of Beijing Normal University and Tianhe 2JK at the Beijing Computational Science Research Center(CSRC).
\end{acknowledgments}

 \bibliography{ref.bib}
 \clearpage
\appendix

\section{Convergence $S_2$ as a function of projection power $m$}
\label{S2mconverge}
\begin{figure}[h]
     \centering
     \includegraphics[width=0.45\textwidth]{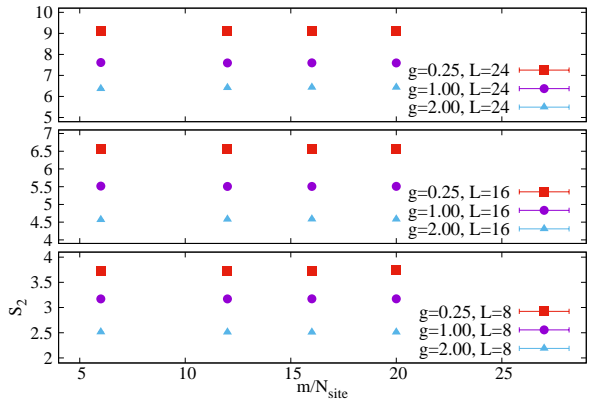}
	\caption{The half-system second R$\Acute{e}$nyi entanglement entropy $S_2(L)$ for the bilayer Heisenberg model at $g=0.25,1.0,2.0$ 
	as a function of projection power $m$ with system sizes $L=8, 16, 24$, respectively. $m=20N_{s}$ is big enough for convergence to ground state value. }
     \label{mtest1}
 \end{figure}

In Fig. \ref{mtest1}, we show the convergence of the second R\'enyi entanglement entropy $S_2(L)$ for the bilayer Heisenberg model at different $g$ 
as functions of the projection power $m/N_s$ for several system sizes. It is evident that $m/N_s=20$ is large enough to probe for the ground state 
properties.

\section{Inverse temperature scaling}
\label{beta_gs}
\begin{figure}[h]
     \centering
     \includegraphics[width=0.45\textwidth]{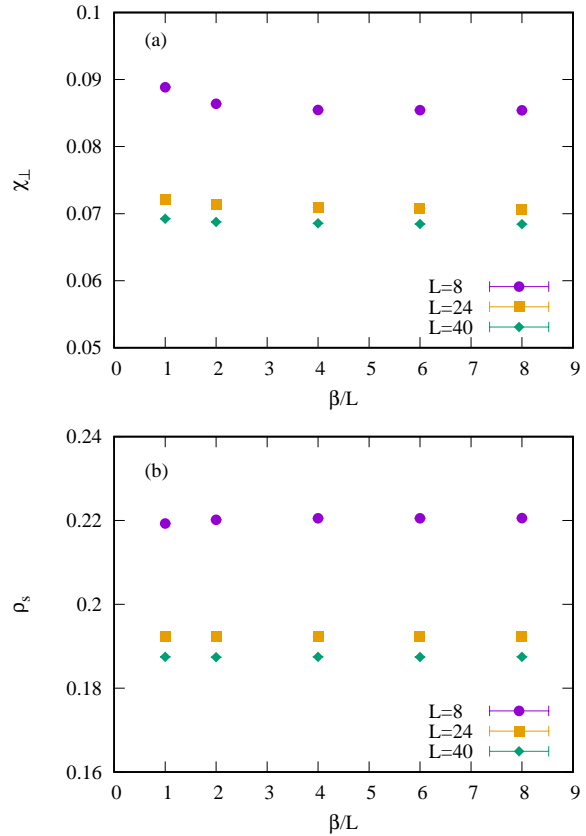}
	 \caption{(a) $\chi_{\perp}(L)$ and (b)$\rho_s(L)$ of the Heisenberg model versus $\beta/L$ with the system sizes $L=8,24,40$.
        As shown, $\beta=4L$ is large enough  to converge  the finite temperature values of $\chi_\perp(L)$ and $\rho_s(L)$ to the ground state values.
        }
     \label{betatest1}
 \end{figure}

\begin{figure}[h]
     \centering
     \includegraphics[width=0.45\textwidth]{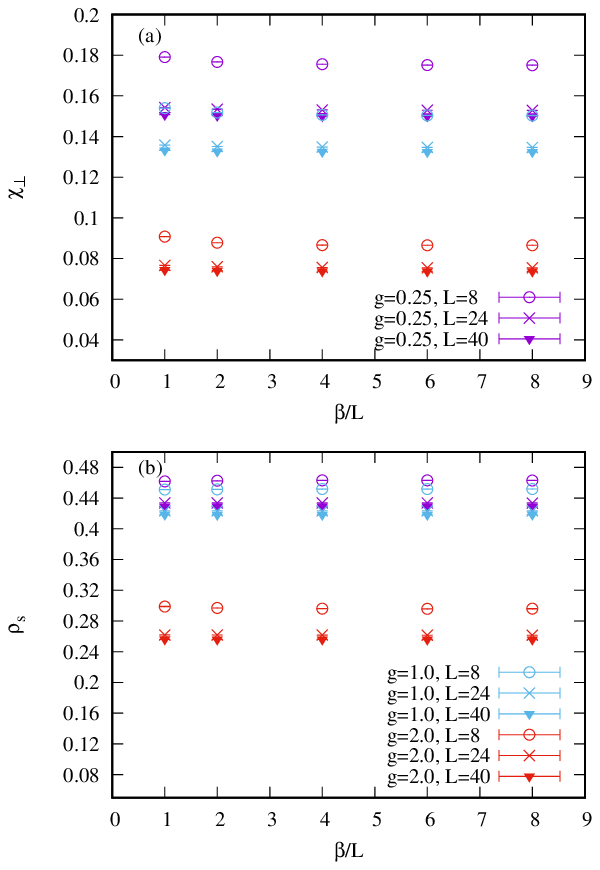}
	 \caption{(a) $\chi_{\perp}(L)$ and (b) $\rho_s(L)$ of the bilayer Heisenberg model versus $\beta/L$ with the system sizes $L=8,24,40$ at several $g$.
        As shown, $\beta=6L$ is large enough  to converge  the finite temperature values of $\chi_\perp(L)$ and $\rho_s(L)$ to the ground state values.
        }
     \label{betatest2}
 \end{figure}

To find out the sufficient large inverse temperature $\beta$ for simulating the properties of the ground states, we 
plot $\chi_{\perp}(L)$ and $\rho_{s}(L)$ versus $\beta/L$ for the Heisenberg model and the bilayer Heisenberg model at different $g$ in Fig.\ref{betatest1} and Fig.\ref{betatest2}, respectively.
It is evident that $\beta=4L$ is large enough to converge $\chi_\perp(L)$ and $\rho_s(L)$ to their ground state values for the Heisenberg model, while 
$\beta=6L$ is large enough to converge $\chi_\perp(L)$ and $\rho_s(L)$ to their ground state values for the bilayer Heisenberg model.

\section{Fit parameters for $\chi_\perp(L)$ and $\rho_s(L)$ of the bilayer Heisenberg model}
\label{app_ab}
Table \ref{tab:ab} lists the fit parameters $a_{1}, a_{2},b_{1},b_{2}$ from  Eq.(\ref{chi_fss}) and Eq.(\ref{rhos_fss}) for the bilayer AF Heisenberg model 
at $g=0.25, 1$, and 2.

\begin{ruledtabular}
 \begin{table}[thb]
	 \caption{The fit parameters $a_1, a_2$ from Eq. (\ref{rhos_fss}) and $b_1, b_2$ from Eq. (\ref{chi_fss}) for the bilayer Heisenberg model at
	 various $g$. }
     \begin{tabular}{c|c|c|c|c}
         g  &   $a_{1}$         & $a_{2}$     &$b_{1}$ & $b_{2}$\\
    \hline
       0.25 &  0.59(2)   & 1.1(2)     & 0.76(2)             &  6.7(2)          \\
    \hline
          1 &  0.619(9)  & 1.26(6)     & 0.72(1)             &  4.54(9)           \\
    \hline
          2 &  1.05(2)   & 3.5(2)     & 1.20(2)             &  4.1(2)            \\
     \end{tabular}
     \label{tab:ab}
 \end{table}
\end{ruledtabular}

\section{Fit parameters vary against minimum size $L_{\rm min}$ used in the fit}
\label{lmin}
 Tables \ref{tab:g0.25}, \ref{tab:g1}, and \ref{tab:g2} show fitted coefficients of the area law, the logarithmic term, and the constant vary against 
 minimum size $L_{\rm min}$ for the bilayer Heisenberg model at different $g$.

  \begin{ruledtabular}
 \begin{table}[thb]
  \caption{Coefficients of the area law, the logarithmic term, and the constant with $g=0.25$.}
    \begin{tabular}{c|c|c|c|c}
         $L_{\rm min}$ &  a           & $N_G/2$    &$\gamma_{\rm ord}$ &$\chi_r^2$/P-value \\   
    \hline
                8  &  0.275(1)   & 0.97(1)   & 0.75(3)                  & 1.65/0.13   \\
    \hline
                10 &  0.273(1)   & 1.00(3)    & 0.74(5)                  &1.60/0.16 \\
   \hline
                12 &  0.271(2)   & 1.05(4)    &    -0.72(8)               & 1.09/0.36 \\
     \end{tabular}
     \label{tab:g0.25}
 \end{table}
 \end{ruledtabular}
 
  \begin{ruledtabular}
 \begin{table}[thb]
  \caption{Coefficients of the area law, the logarithmic term, and the constant with $g=1.0$.}
    \begin{tabular}{c|c|c|c|c}
         $L_{\rm min}$ &  a           & $N_G/2$    &$\gamma_{\rm ord}$ &$\chi_r^2$/P-value \\   
    \hline
                8  &  0.211(1)  & 1.03(2)   &  0.73(3)                & 0.84/0.54   \\
    \hline
                10 &  0.210(1)   & 1.05(3)    & 0.71(5)                  &1.60/0.16 \\
   \hline
                12 &  0.210(2)   & 1.06(4)    &   0.70(7)                & 0.81/0.52 \\
     \end{tabular}
     \label{tab:g1}
 \end{table}
 \end{ruledtabular}

  \begin{ruledtabular}
 \begin{table}[thb]
  \caption{Coefficients of the area law, the logarithmic term, and the constant with $g=2.0$.}
    \begin{tabular}{c|c|c|c|c}
         $L_{\rm min}$ &  a           & $N_G/2$    &$\gamma_{\rm ord}$ &$\chi_r^2$/P-value \\   
    \hline
                8  &  0.183(1)   & 1.04(2)   &    0.72(4)               & 0.7/0.65   \\
    \hline
                10 &  0.184(1)   & 1.02(3)    & 0.76(5)                  &0.45/0.81 \\
   \hline
                12 &  0.185(2)   & 0.99(4)    &   0.74(9)                & 0.38/0.82 \\
     \end{tabular}
     \label{tab:g2}
 \end{table}
 \end{ruledtabular}
 
\end{CJK*}
\end{document}